       \documentstyle [12pt]   {article}
 \evensidemargin\oddsidemargin
\begin{document}
\count0 = 1
 \title{ Irreversibility, Information and Randomness\\
   in  Quantum Measurements }
\author{ S.N.Mayburov \\
Lebedev Inst. of Physics\\
Leninsky Prospect 53, Moscow, Russia, 117924\\
E-mail :\quad   mayburov@sci.lebedev.ru}
\date {}
\maketitle

\begin{abstract}
Irreversibility in quantum measurements is considered from the
point of
 quantum information theory. For that purpose the
information transfer between the measured object $\Omega$ and
measuring system $\cal O$ is analyzed.
  It's found that due to the principal constraints of quantum-mechanical
  origin,
  the  information about the purity of $\Omega$ state isn't
transferred to $\cal O$ during the measurement of arbitrary
$\Omega$ observable $V$.
 Consequently $\cal O$ can't discriminate
the pure and mixed $\Omega$ ensembles with the same $\bar V$. As
the result,  the  random outcomes should be detected by $O$ in $V$
measurement for $\Omega$ pure ensemble of $V$ eigenstate
superposition.
 It's shown that the outcome
probabilties obey to
Born rule. The influence of $\cal O$ decoherence by its
environment is studied,
however the account of its effects doesn't change
these results principally.\\

\end{abstract}

\section  {Introduction}


 Quantum mechanics (QM) after more than 90 years of its
development  achieved the tremendous success in the description
of nature. However, its foundations are still disputed extensively
 and seems to contain some unsettled questions $\cite {1,2}$.
 The most famous and oldest  of them is
   State Collapse or  Measurement
   problem $\cite {1,2,5}$.
In its essence, the experimental measurement of pure quantum
state  parameters  result in the random outcomes with
probabilities described by Born rule, such situation seems to
contradict to fundamental QM linearity \cite {2,5}. This effect
called also the irreversibilty of 2-nd kind supposes that
$\Omega$ interaction with measuring device is nonunitary and
irreversible, whereas in all other situations quantum evolution
is known to be unitary.

In our paper this problem
  will be considered mainly
within the framework of information theory $\cite {1,5}$.
 Really, the  measurement of physical parameters, characterizing an
arbitrary object $\Omega$, includes the transfer of information about
 $\Omega$ state  to the information system  $\cal O$  (Observer) which processes
 and memorizes it $\cite {1,5}$. Correspondingly, in information theory
 any measuring system (MS)
can be considered  as the information channel connecting $\Omega$
and $\cal {O}$ \cite {5,6,7}. Such approach, in principle, can
have important implications for the
 theory of quantum measurements.
In particular, if some constraints on the information
 transfer via such channel exist, they can influence the
   information available for  $\cal O$ during the measurement of $\Omega$
state and distort the measurement results.
Quantum information  studies have shown earlier that such
 quantum constraints
 are really exist in typical information channels and
  define the  channel capacity $\cite {6,7}$.
   However, until now the influence of this and other information-theoretical effects on the outcomes of
 quantum measurements wasn't analyzed thoroughly.
 Basing on these premises, in our previous papers the measurement
 of $\Omega$ observable $V$ by $\cal O$ was analyzed for simple MS model $\cite {8,9}$.
  It was shown that under simple assumptions an arbitrary $\Omega$ observable $V$ can be effectively measured by $\cal O$,
   whereas the information characterizing  the purity of $\Omega$
  state isn't transferred to $\cal O$ at all.
As the result,  $\cal O$ can't
discriminate the pure and mixed
 $\Omega$ ensembles with the same $\bar V$.
 For individual events such information
losses result in the appearance  of random $\cal O$ outcomes in
the measurement of $V$ eigenstate superposition $\cite {9,10}$.

Here the information-theoretical approach of system
self-description will be applied
 to the analysis of information acquisition by $\cal O$  and  signal recognition $\cite {5,3}$.
 The formalism of restrictive maps in MS Hilbert space will be used
 for the calculation of information transfer from $\Omega$ to $\cal O$
 during the measurement \cite {4}.
 Its application  permits to drop the  majority of
 $ad \, hoc$ assumptions used earlier in our papers.
  It will be shown also
that beside the information losses in MS, the initial $\Omega$
information about purity is by itself principally incomplete and
so insufficient  for pure/mixed state discrimination in a single
event.
 Basing on these  results, the model-independent restriction on
state purity information will be obtained. It supposes that the
randomness is generic for standard quantum measurements and so QM
Reduction postulate  is excessive in QM formalism \cite {10}. The
considered effect isn't related directly to  MS decoherence by
its environment, yet its account performed in our paper
  enlarges the considered information losses
   and stabilizes the final $\cal O$ states $\cite {12}$.

Plainly, the most detailed  measurement study should  include the
analysis  of individual events. In quantum case the individual
quantum states are exploited for that purpose $\cite {1,2}$.
 We shall consider here only
 such individual states which can be prepared 'event by event' by
 experimentalist in the idealized scheme of state preparation \cite {2}.
 For the
finite-dimensional system $S$ they are the pure states (rays) in
$S$ Hilbert space $\cal H$.
The statistical or ensemble states
 are described by the density matrix $\rho$,
i.e. the positive trace one operator on $\cal H$. If the
individual procedure of ensemble preparation is used, such
ensemble
can admit more detailed description in form of gemenge \cite {1},
its example is given below. Only those QM observables , which
correspond to the linear, self-adjoint operators, will be used in
our formalism;
 POV generalization of  QM observables is
unimportant for that.

In information theory
   the outcome of $\Omega$ measurement
   by $\cal O$ in event $n$
 is described by the array of real or discrete parameters called
 the information pattern (IP)
$J(n)=\{e^1,... ,e^l\}$ \cite {11}. For example, if $\cal O$
measures the spin projection and momentum of some particle, then
the resulting IP is: $J(n)= \{S_z, P_x, P_y, P_z\}$.
 The  difference between two
 signals for $\cal O$ is reflected by the difference
 of their IPs, its typical measure is:
 $dJ=\sum |e^i_1-e^i_2|$ \cite {11}.
 Usually, it is
admitted implicitly that $\cal O$ physical structure permits to
perform such operations, which result in the recognition of
incoming signals, yet for quantum $\cal O$ models studied here
this is the additional assumption.

\section                   {Model of Quantum Measurements}

 In our model MS consists of measured object
   $S$, detector $D$
 and the information  system  $O$.
 $D$ and  $O$ supposedly can be treated as the quantum objects and
   MS, as the whole, is
described by the quantum state
 $\Psi_{MS}(t)$ relative to  some external  observer or reference frame (RF) $O'$.
$S$ is taken to be the particle with the spin $\frac {1}{2}$ and
its $z$ projection $S_z$ is measured. Its $u$, $d$ eigenstates are
 denoted $|S_{1,2}\rangle$, so that
  their superposition  has the form:
\begin {equation}
 \Phi_S =a_1|S_1\rangle+a_2|S_2\rangle
  \label {AA99}
\end {equation}
 To compare the measurements of similar pure and mixed $S$
 ensembles,
  $|S_i\rangle$   ensemble with the same $\bar{S}_z$ will be used.
 From its preparation such mixed $S$ ensemble can be described as the gemenge:
 \begin {equation}
W^s=\{|S_i\rangle, P_i \}                          \label {AA33}
 \end {equation}
 where ${P}_i$ $=|a_i|^2$ are
  $|S_i\rangle$ ensemble probabilities \cite {1};
  its  density matrix denoted $\rho_s$.

Normally, $D$ amplifies $S$ amplitude to the level accessible for
$O$ processing, here for the simplicity it just doubles it. $D$
pure states are  described by Dirac vectors in two-dimensional
%
%
  Hilbert space $\cal {H}_D$. Its basis is  constituted
by   $|D_{1,2}\rangle$ eigenstates of $\Lambda$ 'pointer' observable
with eigenvalues $\lambda_{1,2}$.
 The initial $D$ state is:
\begin {equation}
    |D_0\rangle=\frac{1}{\sqrt{2}}({|D_1\rangle+|D_2\rangle}) \label {kkk}
\end {equation}
 It is supposed that  $S$, $D$ interaction starts at
$t_0$ and finishes effectively at some $t_1=t_0+\tau$. For Zurek
$S, D$ hamiltonian $H_{S,D}$ with the parameters tuned optimally
for given $\tau$ value,
the  measurement of
 $S$ eigenstate $|S_{1,2}\rangle$ induces the final product state \cite {12}:
\begin {equation}
 \Psi^{1,2}_C=|S_{1,2}\rangle |D_{1,2}\rangle \label {A44}
 \end {equation}
 From the linearity of Schroedinger evolution it follows
  that for $a_{1,2} \ne 0$ such $S, D$ interaction
  will result in  $S,\, D$  entangled final  state $ \cite {12}$:
\begin {equation}
   \Psi_{S,D} = \sum_{i=1}^2 a_i|S_i\rangle|D_i\rangle
                                   \label {AA2}
\end {equation}
for initial $|D_0\rangle$, $\Phi_S$.

If $a_{1,2}\ne 0$, then $D$ separate state $\Delta_D$ also can be
formally defined, however, due to $S, D$ entanglement,
 $D$ and $S$ properties can't be completely factorized.
It admitted usually that $\Delta_D$ coincides with $D$ reduced
state, i.e. the partial trace over 'external' DFs, which for
$\Psi_{S,D}$  is equal to:
\begin {equation}
   R^e_{D} = \sum_{i=1}^2 |a_i|^2 |D_i\rangle\langle D_i|
                                   \label {A222}
\end {equation}
in terms of density matrixes. This expression is rather obvious
for $D$ statistical state, but
 for the individual state such definition seems to be
controversial, first of all, because $R^e_D$ isn't the pure state
in $\cal H_D$. This formal difficulty, in fact, is unimportant,
because in all calculations $\Psi_{S,D}$ can be used in place of
$\Delta_D$. The  proper ansatz  for the separate states will be
discussed below, until then it will be no need to use it.

Our model of information system $O$ assumes that it's  the
quantum object and analogously to $D$,  its states are
  defined on two-dimensional Hilbert space $\cal H_O$.
Before the measurement starts $O$ initial state is equal to:
\begin {equation}
|O_0\rangle=\frac{1}{\sqrt{2}}(|O_1\rangle+|O_2\rangle)
  \label{ooo}
\end {equation}
 where
$|O_{1,2}\rangle$ are eigenstates of $O$ 'internal pointer'
observable $Q$ with eigenvalues $q_{1,2}$.
  $D$, $O$ interaction starts at some
 $t_2>t_1$
  and finishes at $t_3=t_2+\tau$, during this time interval
 the information about $D$ state is transferred to $O$.
  $D$, $O$ interaction is also described by Zurek hamiltonian
$H_{D,O}$ with the same parameters as $H_{S,D}$. Under this conditions  the
 incoming $\Psi_{S,D}$, $|O_0\rangle$ states  will evolve into:
\begin {equation}
   \Psi_{MS} = \sum_{i=1}^2 a_i|S_i\rangle|D_i\rangle |O_i\rangle
                                   \label {A70}
\end {equation}
relative to external RF $O'$. Such triple decomposition is unique
and in this sense defines $|O_{1,2}\rangle $ as the preferred
basis (PB) of $O$ states $\cite {13}$. In principle,
 $O$ can include other degrees of freedom (DFs) which participate
 in information processing, yet we shall suppose that during $D, O$ interaction
they don't interact with $O$ DFs described by  $\cal H_O$.
 The effects of MS decoherence by the environment
 will be considered separately, because our main results don't depend
on them directly.

\section { Measurements and Information Acquisition }


In information theory the most general and mathematically
powerful approach to the measurements is introduced by the
formalism of system self-description \cite {5,3}.
 To illustrate its meaning,
let's consider some information system $\cal O$, which measures
the parameters of arbitrary object $\Omega$. Then, $\cal O$ can
be  considered formally as
 the subsystem of 'complete' system $\Xi=\{\Omega$,$\cal O$$\}$,
 the set of $\Xi$  states  denoted $N_T$,
so that $N_O \subseteq N_T$ where $N_O$ is the set of $\cal O$
states.
 When $\Omega$, $\cal O$  interaction is finished, $\Xi$ will be
  in some final state $\Gamma$, which for the effective measuring
set-up would be  correlated  with the initial $\Omega$ state
$\varphi_{in}$.
  Therefore, the  measurement of $\Omega$  state  by  $\cal O$ in this approach
 is equivalent to $\Gamma$ measurement by $\Xi$ subsystem $\cal O$.
 Hence it can be described as the mapping of $\Xi$ set $N_T$
 to its subset $N_O$, i.e. to itself, so such process can be
called $\Xi$ measurement from inside $\cite {4}$. The restrictive
map $M_O \Gamma \to R$ describes the restriction of $\Xi$ state
to $\cal O$, by the slight abuse of definitions, $R$ called also
$\cal O$ restricted state. The inverse map $M_O^{-}R \to \Gamma$
is called the inference map.
 In practice, the information acquisition by $\cal O$
always correlates with the change of  its internal state,
correspondingly, in our approach $R$ should be correlated with
$\Gamma$.
 For example, if $\cal O$ is the  atom, $R$ can be the
state of its electronic shells, their excitations would 'record'
the incoming $\Omega$ signals.


 The important property of $\Xi$ restrictions
 is formulated by Breuer Theorem: if for two arbitrary $\Xi$
states $\Gamma,\Gamma'$ their restrictions $R, R'$ coincide, then
for $\cal O$ this $\Xi$ states are indistinguishable $\cite {3}$.
It follows that for any nontrivial
 $\Xi, \cal O$ at least one pair of such $\Xi$ states  exist.
%
  For classical systems the  incompleteness of $\Xi$
   description by $\cal O$ has the
  obvious reason:  $\cal O$ is only the part of $\Xi$, but it should
describe its own  state $R$ plus the state of $\Xi$ 'residual',
 hence $N_T$ mapping to $N_O$ can't be unambiguous $\cite {5,3}$.
 Constructing the quantum self-description
formalism, we shall follow standard QM axiomatic, if no direct
contradictions to it would appear.
$R$ is $\cal O$ internal state, so let's suppose for the start
that it defined on $\cal H_O$ only, below this assumption will be
reconsidered.

It's reasonable to study first the restrictions of $\Xi$
statistical states $\Gamma_{st}$ which  derivation is more simple.
 The expectation values of all $\cal O$ observables
should be the same both for given $\Gamma_{st}$ and its
 $\cal O$ restriction $R_{st}$ \cite {10,4}.
 Then, from QM
 correspondence between the set of such expectation values
 and statistical states it follows that the only consistent solution
 for $R_{st}$ is  the
partial trace of $\Xi$ state over $\Omega$ DFs, i.e.
is $\cal O$ reduced state.
 For our MS model $\Omega$ formally corresponds not to $S$, but
to $S, D$ subsystem,  the statistical restriction corresponding to
  MS  state of (\ref{A70}) is equal to:
\begin {equation}
   R_{st}=Tr_{\Omega} \,  {\rho}_{MS}=\sum |a_i|^2|O_i\rangle\langle O_i|
      \label {AB4}
\end {equation}
where $\rho_{MS}=|\Psi_{MS}\rangle \langle \Psi_{MS}|$.

%
Let's start the study of individual $\Xi$ restrictions from the
situations when
 $\Xi$ final state is the tensor product of $\Omega$, $\cal O$ states.
 For our MS model they appear in the
measurements of $S_z$ eigenstates $|S_i\rangle$. In this case,
the final MS state is equal to:
\begin {equation}
\Psi_i=|s_i\rangle \otimes |D_i\rangle  \otimes |O_i\rangle
   \label {ZZ1}
\end {equation}
 Plainly, due to the factorization
  of $S, D$ and $O$  states,
   defined  on their own Hilbert spaces,
 $\Psi_i$ restriction  to $O$ is  given by: $\xi_i=|O_i\rangle$.
Really, $\Psi_i$ is $Q$ eigenstate with eigenvalue $q_i$, which is
MS real (objective) property, thus $R_O$ possesses the same real
property. Yet it means that $R_O$ is $Q$ eigenstate with the same
eigenvalue, but the only such $O$ state is $|O_i\rangle$.
Since $Q$ eigenvalues $q_i$ are $O$
 real properties $\cite {1}$,  the difference
 between
the restricted states $ |O_i\rangle $ is also objective.
Therefore, it's plausible
 to admit that in such measurement from inside $O$
 can identify them as the different states
characterized by IP  $J^O=q_{i}$, it can be called the minimal
recognition assumption (MRA).
Due to such unambiguous correspondence with particular IPs,
$|O_{1,2}\rangle$ constitute the 'recognition' basis, the
comparison with it
 will help to derive the measurement outcomes for other
MS states.


Consider now
the individual measurements of  $S$ mixed ensemble $W^s$ of (\ref
{AA33}). By preparation, this is probabilistic mixture of
$|S_{1,2}\rangle$ states, for each of them $S, D$
 and $D, O$
interactions  results in appearance of orthogonal MS states
$\Psi_i$.
Thus,  such MS ensemble is described by the gemenge
$W^{MS}= \{\Psi_i, P_i\}$.
 The corresponding  individual MS state is random, i.e.
 it can change from event to event:
\begin {equation}
      \Upsilon_{MS}= \Psi_1 \, .or.\, \Psi_2
  \label {A45}
\end {equation}
where the frequencies of $\Psi_{1,2}$ appearance
 are described by the same
 probabilities $ P_{1,2}$.
  $\Psi_{1,2}$ restrictions were
 obtained above, so $O$ restriction of such random MS state is equal to:
\begin {equation}
              R_{mix}= |O_1\rangle \, .or.\, |O_2\rangle  \label {UV}
\end {equation}
Each $|O_i\rangle$ appears with the corresponding probability
${P}_{\it i}$, so that the ensemble of $O$ restricted states
described by the  gemenge
 $W^O=\{|O_i\rangle, {P}_{\it i}  \}$
with density matrix:
\begin {equation}
              \rho_{o}= \sum P_i|O_i\rangle \langle O_i|  \label {UW}
\end {equation}

For nonfactorized individual $\Xi$ states Breuer assumed phenomenologically
 that, analogously to the statistical states,
 their restrictions  are equal to
  $\cal O$ reduced states $\cite {3,4}$.
For  $\Psi_{MS}$  of (\ref {A70}) it gives:
\begin {equation}
   R_B=Tr_{\Omega} \, {\rho}_{MS}=\sum |a_i|^2|O_i\rangle\langle O_i|
      \label {AA4}
\end {equation}
Plainly, this ansatz excludes beforehand any
kind of stochastic behavior for  MS restriction.
The resulting  $R_B $ differs from
 $R_{mix}$  of ({\ref{UV}) which describes the restriction
of corresponding mixed MS ensemble $W^{MS}$.
It supposes  that, in principle, $O$ can discriminate the
individual pure/mixed MS states 'from inside',
 because the condition of Breuer theorem is violated.
 Yet it will be shown below that the  analysis
  of individual measurements   permits to derive the
    MS restrictions  to $O$ unambiguously
   without any $ad \, hoc$ assumptions,
yet  these results will disagree with the former conclusion.
   Note also that even for this simple ansatz the inference map
$M_O^-$ is ambiguous: all MS states  of (\ref {A70}) with the same
$|a_{1,2}|$ has the same restriction $R_B$ of (\ref {AA4}), so
it's not possible, in principle, to choose just one of them from
the knowledge of $R_B$ only.

\section {Discrimination of Individual States}

 As was  shown, the measurement of $S_z$ eigenstates $|S_{1,2}\rangle$
 produces final MS states $\Psi_{1,2}$,
 which $O$ restrictions $\xi_{1,2}$ are  equal to $|O_{1,2}\rangle$;
 MRA claims that they are identified by $O$  as IP $J^O=q^O_{1,2}$.
  Let's calculate in this framework   $O$ restricted state
  which appears in MS measurement   of $|S_{1,2}\rangle$ superposition  (\ref
  {AA99}).
In this case MS final state of (\ref {A70}) $\Psi_{MS} \ne \Psi_{1,2}$,
but by itself, the formal
 difference of two MS individual states
is the necessary but not sufficient condition for their
discrimination by $O$. In addition, such MS measurement from
inside should permit $O$ to detect the difference between the
restrictions of those MS states to $O$.
Previously, in MRA ansatz $J^O_i$ was formally expressed as
$e^1=q_i$ but, in principle,
 it can include  more parameters $e^2,...,e^m$,
 whose values are identical for  $\xi_{1,2}$.
According to Boolean logic, if for $O$ MS restriction $\xi_s$
differs from $\xi_{1,2}$, then $\xi_s$ can be identified by $O$ in
the event of measurement as the different set of real  parameters,
i.e. IP $J^O_s \neq J^O_{1,2}$.
 Therefore, $J^O_s, J^O_{1,2}$ should  include  at least one parameter
 $e^j$,  which value $g_0$ for  $\xi_s$
is different from its values $g_{1,2}$ for $\xi_{1,2}$.
%
In QM framework such $e^j$ should be some MS observable $G$ to
which corresponds  the linear, self-adjoint operator $\hat G$. In
this case $\xi_s, \xi_{1,2}$  will be $G$ eigenstates with the
eigenvalues $g_{0,1,2}$; so $O$ would discriminate $\xi_s$ from
$\xi_{1,2}$, if $g_0\ne g_{1,2}$.
 It was supposed earlier that MS restrictions
to $O$ are defined on $\cal H_O$, so it follows that $G$ should
belong to the set (algebra) of $O$ observables $\cal U_O$.
 In  our MS model $\cal U_O$  is
equivalent to observable algebra of spin$\frac{1}{2}$ object, so
any $O$ nontrivial observable can be expressed as $\cite 2$:
\begin {equation}
  A=d_0Q+d_1Q^x+d_2Q^y     \label {QZ}
\end {equation}
 where arbitrary real $d_i$
coefficients are normalized to $\sum d^2_i=1$.
  $O$ observables $Q^{x,y}$ are conjugated to $Q$ and obey
the standard commutation relations: $[Q,Q^{x,y}]=i\hbar\beta
Q^{y,x}$ where $\beta=1$ for $Q^x$ commutator, $\beta=-1$ for
$Q^y$ one. $\xi_{1,2}$ exhaust the spectra of $Q$ eigenstates and
so $G \ne Q$. In the same time, considering the equation
 $\hat{A}\xi_i=v_i\xi_i$ for real $v_i$, it follows that
 $\xi_i$
can't be the eigenstate of any other $A \ne Q$, hence there is no
$O$ observable $G$ which can satisfy to all our demands
simultaneously.


Thus, only $\xi_{1,2}$  states can be unambiguously
  discriminated by $O$ in such MS measurement from inside,
 there is no IP $J^O_s \ne J^O_{1,2}$
which can be correctly ascribed to $\xi_s$. Since any alternative
outcomes for MS measurement by $O$ are supposedly impossible in
our formalism,
 in particular, 'undefined' or 'incomparable' outcome, the only consistent
$J^O_s$ ansatz  is equal to:
$$
           J^O_s= q_1 \, .or. \,q_2.
$$
As the result, $O$ can't distinguish $\xi_s$ and $\xi_i$ states
and $O$ restriction  of $\Psi_{MS}$ of (\ref{A70})  is equivalent
to:
 \begin {equation}
         \xi_s = \xi_1 \, .or.\, \xi_2.   \label {B3}
\end {equation}
 i.e. it's equivalent to $R_{mix}$ of (\ref{UV}).
 It suppose that
 MS restrictive  map $M_O$ is stochastic,
and  because of it, the corresponding inference map $M_O^-$ is
ambiguous.

%

 Summing up, it means that the ensemble of $O$ restricted states $\xi_s$ is  described by
 the gemenge $W^a=\{|O_i\rangle, {P'}_{i} \}$ where probabilities $P'_i$ should be
calculated. Remind that $W^a$ statistical state $R_{st}$ is
described by density matrix (\ref {AB4}), from that the relation
$P'_i=|a_i|^2$ follows. Thus the probabilities in such subjective
gemenge obey to Born rule.
Note that  Born rule for outcome probabilities isn't self-obvious
in QM formalism, it should be independently derived in any new
theory of measurement $\cite {1,14}$.


\section {Measurement Correlations and Joint Observables}

In our calculations it was supposed that  $O$ restrictions of MS
states can be discriminated by $O$ observables only which seems
quite reasonable.
Yet to be safe, let's relax this condition and check full MS
observable algebra in search of observables
 which can  discriminate the pure and mixed MS ensembles described above.
As follows from the properties of statistical restrictions
 considered in sect. 3,
 if the restriction $\xi_s$ of some state $\Psi_a$
is the eigenstate of some observable $\Lambda_a$, then $\Psi_a$ is
also $\Lambda_a$ eigenstate \cite {Em4}.
In this framework MS states can be  used in search of suitable
observable $G$,  for which the following relations should
 be fulfilled simultaneously:
\begin {eqnarray}
   \hat{G}\Psi_{MS}=g_0\Psi_{MS}\nonumber\\
    \hat{G}\Psi_{1,2}=g_{1,2}\Psi_{1,2}\label {ZZX}
\end {eqnarray}
for $g_0 \ne g_{1,2}$. However, in our MS model $\Psi_{MS}= \sum a_i \Psi_i$,
and the substitution of second equation into the first one
gives: $g_0=g_1=g_2$.
 Hence no MS observable $G$ possesses the different
eigenvalues for $\Psi_{MS}$ of (\ref {A70}) and $\Psi_{1,2}$ of
(\ref {ZZ1}),  so even all MS observables
 would be available  for $O$ measurement from inside it will not
permit $O$ to discriminate such MS states. Only MS observables
corresponding
 to  the nonlinear operators  can reveal
 the  difference between $\Psi_{MS}$ and $\Psi_{1,2}$ restrictions to $O$, but their
measurability  contradicts to standard QM axiomatic. Note that if
only $O$ observables $G$ are considered, then our previous
results of (\ref {QZ}) are reproduced by this ansatz.


These calculations for the measurement from inside are applicable
to arbitrary MS observable, let's compare them with its
measurement
by some external RF $O'$.
It's argued often that if it can be shown that MS
 final state after $S$ measurement is pure for external RF $O'$,
  then it excludes the possibility
   of its random outcomes for $O$.
   As was shown, this supposedly is
    untrue for MS measurement from inside,
    yet such reasoning results in frequent confusions,
    so it's instructive to consider it here.
For MS final states the difference between
their pure and mixed ensembles with the same $\bar {S}_z$
 can be revealed by interference term (IT) observable
 which are the joint $S, D$, $O$  observables $\cite 1$.
 Such IT can't be measured by $O$ 'from inside' for  described MS
set-up, which tuned to the optimal $S_z$ measurement.
  IT general ansatz  is rather complicated,  here only
  symmetric IT will be exploited:
\begin {equation}
B= |O_1\rangle \langle O_2| D_1\rangle \langle
D_2||S_1\rangle\langle S_2|+
    |O_2\rangle \langle O_1| D_2\rangle \langle
    D_1||S_2\rangle \langle S_1|
    \label {AA5}
\end {equation}
Being measured by external $O'$ via its interaction with $S, D$,
$O$, for arbitrary $B$ it gives $\bar{B}=0$ for
 $\Psi_{1,2}$ probabilistic mixture $W^{MS}$, but  for some MS states $\Psi_{MS}$ of (\ref
{A70})  one obtains that $\bar{B}\neq 0$. In particular, for
symmetric $S$ state $\Phi_s$ of (\ref {AA99})
 with $a_{1,2}=\frac{1}{\sqrt{2}}$, the resulting
$\Psi_{MS}$ of (\ref {A70}) is $B$ eigenstate with eigenvalue
$b_1=1$. $B$ possesses also two other eigenvalues $b_{0,2}$, of
them only $b_2=-1$ is important in this case. The probability
 $ P_{ B}( b_{1,2})=.5$
for $\Psi_{1,2}$ mixture with  $\bar{S}_z=0$, the intersection of
its $b$ probability distribution
 with the one for $\Psi_{MS}$ results in their overlap $K_b =.5$.
 Hence, in accordance with our previous conclusions, the pure/mixed MS states
with the same $\bar{S}_z$ can be discriminated even by external
$O'$ only statistically, there is no MS observable which can
discriminate them 'event by event'. It demonstrates that the
operational difference between such MS states is relatively small.
 The properties of other MS ITs  are similar to the symmetric one, but
the difference between pure and mixed MS ensembles is less
pronounced. The joint $S, D$ observables posses the similar
properties, in particular, symmetric  IT $B_{S,D}$ can be
obtained from ( \ref {AA5}), if to remove all $O$ terms. It can
be measured by $O$ via the simultaneous interaction with $S$ and
$D$.


 The proposed theory admits
  that in general the same MS pure state can look stochastic
for $O$ measuring it from inside, but in the same time
 can evolve linearly relative  to some external $O'$.
It was shown earlier that such situation by itself doesn't lead to
any experimentally observed inconsistency for the results of
measurements which can be performed by $O$ and $O'$ $\cite
{9,10}$. The consistent description of this situation can be
given by the formalism of unitary nonequivalent
 representations $\cite {Em4}$. In particular, in our model
 MS restricted states are defined on $\cal H_O$, which is the
 subspace of MS Hilbert space $\cal H_C$.
 Correspondingly,   the transformation of MS states
 between $O$ and $O'$ will be  nonunitary, i.e. there is no unitary
operator $\hat U$, for which
 $\cal H_{C}$=$\hat {U} \cal H_O$$\hat{U}^{-1}$. Yet only if such
 $\hat U$ exists,
 MS state, which is pure for $O'$, would be also with necessity the pure state for $O$ $\cite 2$.

It's well known that the decoherence of pure states by its
environment $E$ is the important effect in quantum measurements
$\cite {12}$. In the simplified calculations its formalism permits
to suppose that MS, $E$ start to interact only at the final stage
of $S$ measurement.
If $D, O$  interact with $E$ only at $t>t_3$, for the typical
decoherence hamiltonian
 it follows that $\Psi_{MS}$ of (\ref {A70}) will evolve into
 MS, $E$ final state:
\begin {equation}
    \Psi_{MS,E}=\sum_{i=1}^2 a|S_i\rangle |D_i\rangle
|O_i\rangle \prod\limits^{N_E}_{j} |E^j_i\rangle \label {RRR}
\end {equation}
where $E^j$ are $E$ elements,
 $N_E$ is $E^j$ total number. If an arbitrary
$O$ pure state $\Psi_O$ is prepared, it will  also decohere in a
very short time into the analogous $|O_i\rangle$ combinations
entangled with $E$. Thus, of all pure $O$ states, only
$|O_i\rangle$ states are stable relative to $E$ decoherence.
Hence it advocates the choice of such states as $O$ preferred
basis, since in such environment $O$ simply can't percept and
memorize any other $O$ pure state during
 any sizable time interval.  $D, O$ decoherence by $E$  makes  the
considered discrimination of pure and mixed final MS states by
external $O'$ quite complicated, but the analysis  of
corresponding ITs show that
 their main properties   don't
change principally \cite {9,10}.

\section {Information Incompleteness and Measurements}

Now we shall demonstrate  that our results for $S_z$ measurement
from inside by $O$ can be obtained avoiding the direct use of
self-description formalism or, at least, its most sophisticated
part. In particular, it will be argued that the incompleteness of
information carried by individual $S$ state stipulates the
randomness in $S_z$ measurement by $O$, and so this effect is in
some sense is objective and observer-independent.
For this study it's worth to have the statistical estimate of state
discrimination in the measurement of particular observable. Such
statistical measure for two finite-dimensional states
$\rho_{1,2}$ and observable $\Lambda$ can be described as the
coincidence rate (overlap) of their $\lambda_i$ eigenvalue
distributions \cite {7}:
\begin {equation}
         K(\Lambda)= \sum\limits_i [w_1 (\lambda_i) w_2
    (\lambda_i)]^{\frac{1}{2}}  \label {YX}
\end {equation}
here $w_{1,2}(\lambda_i)  =Tr \rho_{1,2} \Pi(\lambda_i)$ where
$\Pi(\lambda_i)$
 is the orthogonal projector on $\lambda_i$.
In particular, the difference between the pure and mixed $S$
states is indicated by $S$
 observables, which  expectation value is sensitive to the
presence of the  component interference.
 For the regarded $S$ pure/mixed states with the same $\bar{S}_z$ they are
 $S_{x,y}$ linear forms.
  For example, if  $\frac{a_1}{a_2}$ is real, the maximal distinction
reveals $S_x$ observable, for which
 $|\bar{S}_x|= |a_1||a_2|$ for the
 pure states and $\bar{S}_x=0$ for the mixture.
In this case, their overlap
$$
K(S_x)=1-|a_1||a_2|
$$
 For the arbitrary $a_{1,2}$ the maximal discrimination of pure
and mixed $S$ states gives the expectation value of observable:
\begin {equation}
S_{\gamma}=S_x\cos \gamma +S_y\sin \gamma \label {SSS}
\end {equation}
where $\gamma$ is  $\psi_s$ quantum phase between
$|s_{1,2}\rangle$ components.
 The value of $r_p=2|\bar{S}_{\gamma}|$, which lays between $0$ and $1$,
 can be chosen as  $S$ purity rate.
 These estimates indicate that
 even the incoming  pure and mixed $S$ states with the same $\bar{S}_z$ differ,
 in fact, only statistically with the minimal overlap  $50\% $,
  but not on 'event by event' basis. Hence in such case the purity
 can be measured consistently only for $S$ ensemble and not for
 individual state.
 Let's consider the final state $\Psi_{S,D}$ of (\ref {AA2})
   and corresponding mixed ensemble induced by gemenge $W^s$ of
   (\ref {AA33})
   with the same $\bar{S}_z$. For $D$ pointer observable $\Lambda$
and all $D$ observables conjugated to it,
 their overlap between pure and mixed states
  $K(\Lambda), K(\Lambda_{x,y,\gamma})$ is equal to $1$ for arbitrary $a_{1,2}$; here  $\Lambda_{x,y,\gamma}$
are defined by the analogy with $S$ observables of (\ref {SSS}).
Hence even statistically no information about $S$ state purity is
transferred to $D$ via MS information channel, which is tuned to
the optimal $S_z$ value measurement. The similar results were
obtained for information transfer via quantum channels \cite {7}.

In this framework let's consider  the information content of
individual $S$ state. Plainly, $S_z$ eigenstate $|S_i\rangle$
transfers $1$ bit of information in $S_z$ measurement
corresponding to the choice of two possible $S_z$ values $\pm
\frac{1}{2}$. Correspondingly, the overlap (\ref {YX}) of such
states $K(S_z)=0$. In this vein let's calculate  the amount of
information about $S$ state purity $I_p$ for symmetric $\Phi_s$
with $a_{1,2}=\frac{1}{\sqrt{2}}$.
 As was shown above, the minimal overlap with the corresponding
 $|S_i\rangle$ mixture $W^s$ of (\ref {AA33}) is dispatched by $S_x$ observable and gives $K(S_x)=.5$
for two such ensembles.  In this case one can conclude
 that for such $S$ states the information $I_p$ is described only by 'half-bit' of
information  per event at MS input. Hence even if $O$ in place of
$S_z$ would measure $S_x$, $S$ purity can't be defined in a
single event. The same or lesser $I_p$ value can be obtained for
arbitrary $\Phi_s$ if $O$ knows the phase $\gamma$ of (\ref
{SSS}). If it's unknown for $O$ then $I_p$ at MS input will be
formally two time less, i.e. less than $\frac{1}{4}$ bit. These
results demonstrate that the amount of information about purity
carried by the individual $S$ state is principally insufficient
for discrimination of pure and mixed $S$ ensembles
 on 'event by event' basis, because
the operational difference between the pure and mixed states is
too small for that.  Plainly, no $S$ quantum interaction with
other objects can enlarge its amount \cite {7}.
 For individual states the difference
between such pure and mixed $S$ states is described only by the
observables related to nonlinear operators. For standard QM
observables such purity information can be extracted only from the
simultaneous joint measurements of large ensembles \cite {1,14}.

Let's study how such incomplete information about $S$ purity is
transferred to $D$ and then to $O$ and what is its effect. Remind
that in our MS model after some time
 moment $t_1$ $S$ and $D$ stop to interact, whereas $D$ and $O$ start to interact
at $t_2>t_1$, so that even for arbitrary $H_{D,O}$ interaction
$O$ can measure directly only $D$ observables. Really, at $t>t_2$
the object $S$ can be miles away from $D$ and $O$, in this case
$S$ and $S, D$ observables surely will  be unavailable for $O$
directly. Thus, one can regard MS as the information channel,
which transfers first $S$ signal to $D$, and after that $D$
signal to $O$.
 In MS model with
$H_{D,O}$ exploited here  the measurement of $S_z$ eigenstates
$|s_i\rangle$ induces factorized MS  state $\Psi_i$
 of (\ref {ZZ1}); in this case $O$ interacts with $D$ separate state
 $\Delta_D=|D_{1,2}\rangle$ and, as was shown, $O$ percepts it as IPs
$J^O_{1,2}=q_{1,2}$.


Consider now $S_z$ measurement for the incoming $|S_i\rangle$
probabilistic mixture (gemenge) $W^s$ of (\ref {AA33}) with some
$\bar {S}_z$. When $S, D$ interaction is finished at $t_1$, then
$S,D$ ensemble becomes
 the mixture (gemenge) of  $\Psi^{1,2}_C$  of (\ref {A44})
  with $\bar{Q}=\bar{S}_z$.
 As was shown above, at $t>t_3$ when $D$ measurement by $O$ is finished,
 its result will be percepted by $O$ as $J^O_1$ or $J^O_2$ with
the probabilities $P_{1,2}$. In the same vein, consider the
possible outcome for pure $S$ state $\Phi_s$ of (1) with the same
$\bar{S}_z$. In this  framework at the final stage of $S_z$
measurement, $D$ separate state $\Delta_D$ interacts with $O$,
which results in appearance of some $O$ IP $J^O_s$, which can
either coincide with one of the basic $O$ IPs $J^O_{1,2}$ or
differ from them. As was shown in sect. 5,
 $\Psi_{S,D}$ of (\ref {AA2}) and $\Psi^{1,2}_C$  can't be the
  nondegenerate eigenstates
of the same $D$ observable $G$. Thus, even if $O$ can measure all
$D$ observables simultaneously, it wouldn't permit $O$ to detect
the difference between such pure and mixed $S$ ensembles; so
 for such pure ensemble $O$
would percept in the individual events IP:
  $J^O_s=J^O_{1} .or. J^O_2$
with probabilities $P_{1,2}$ correspondingly, their values are
defined by $\bar{S}_z$.
Really the opposite result, i.e the observation by $O$ of such
difference would mean that $O$ can measure $D$ observable which
corresponds to the nonlinear operator, but this  contradicts to
 standard QM. The obtained results don't mean that in the pure case the
separate $D$ state $\Delta_D$ of (\ref {A222}) is the objective
 probabilistic mixture of $|D_{1,2}\rangle$,
 rather $\Delta_D$ can be characterized
  as their 'weak' superposition stipulated by the entanglement of
 $S$, $D$ states. In this framework $R^e_D$ of (\ref {A222})  can be regarded
 as the symbolic expression of this difference. Yet the complete description
of $D$ properties is performed only by $S, D$ state as the whole,
so that some of them are described by the nonlocal $S, D$
observables.
 However, no measurement performed on $D$ only
   can reveal the difference of $\Psi_{S,D}$
    from the corresponding $\Psi_C^i$ mixture.
   Only the measurement of some joint $S, D$
observables, like $B^{S,D}$ can reveal it, but also only
statistically \cite {1}.

 These results indicate  that in this set-up $D$ state
 doesn't contain the information about $S$ purity and so
it principally can't be dispatched to $O$. This conclusion
doesn't change even if to admit  that $O$ can measure the joint
$S, D$ observables.
If the information about $S$ purity isn't transferred by $D$, then
$O$ functioning by itself plays the minimal  role in the appearance
of the outcome randomness. The only feature which $O$ should
possess is the proper discrimination of $|D_{1,2}\rangle$ states
as $J^O_{1,2}$.
 Those semi-qualitative arguments aren't
sufficient for the consitent proof of measurement randomness
without exploit of self-description formalism. Yet they evidence
that such randomness can be the consequence of fundamental
information incompleteness of individual quantum states and so
can be regarded as observer-independent.



\section {Discussion}

 The presented calculations, in our opinion, reveal the origin of
  the principal randomness of quantum  measurements.
As follows from our considerations, the structure of QM Observable
Algebra which includes only the linear, self-adjoint operators
 corresponds to Boolean logics of signal
recognition. In our case its operands correspond to IP set
 $\{ J^O\}$ \cite {11}. In particular, it excludes the simultaneous $O$
registration of two opposite outcomes of measurement, which is
the essence of 'Schroedinger cat' paradox. It permits to suppose
that the independent Reduction postulate is unnecessary in QM. It
follows that all the measurement features  can be deduced  from
QM axiom which postulates QM observables to be the linear,
self-adjoint operators and settles the relation between their
eigenstates and the outcomes of corresponding measurements.
 In this approach the randomness in quantum measurements is  related
 to the  incompleteness
 and undecidability aspects of information theory, their studies
 were initiated by notorious  G$\rm \ddot o$del theorem $\cite 5$.
 The considered phenomenon can be called
the subjective collapse of quantum state, because MS as the whole
is in the pure state throughout the measurement relative to
external observer $O'$. Correspondingly, the considered effect
represents also the subjective irreversibilty, induced by the
incompletness of $O$ information about MS state and itself.

 These considerations are closely related to the
question  whether this theory is applicable to  human observer
$O$, in particular, whether in this case IP $J^O$ can describe
 the true $O$ 'impressions' concerned with the measurements' outcomes?
 This is open problem,
but at the microscopic level the human brain, as the dynamical system,
 should obey QM laws as any other object,
  so we don't see any serious reasons to make
the exceptions $\cite {5,9}$. In our model the detection  of
eigenstate $|D_i\rangle$ by $O$ can be associated with the
excitation of some $O$ internal levels. This process is similar
to the excitation of brain molecules during the acquisition of
external signal. In this vein MRA used in our approach looks
reasonable and consistent.
 Note also that in our theory the brain
or any other processor $O$ plays only the passive role of signal
receiver.
 The real effect of information loss which stipulates the outcome
 randomness
occurs 'on the way' when the quantum signal passes
 through MS information channel. Hence the observer's
 consciousness, in principle, can't have any relation to it.

We conclude that standard Schroedinger QM formalism
 together with the information-theoretical considerations
permit to derive the 'subjective' collapse of  pure states without
implementation  of independent Reduction postulate into QM
axiomatic.
 In our approach the main
sources of randomness  are the principal constraints on the
transfer of information in $S\,\to O$
 information channel and the incompleteness of information about $S$ purity
 carried by individual $S$ state.
 This 'lost'  information
characterizes the purity of S state,
 because of its loss, $O$ can't discriminate the pure and mixed $S$ states.
 As the result of this
information incompleteness, the randomness of measurement outcomes
appear, the probabilities of $O$ outcomes obeys to Born rule.





\begin {thebibliography}{0}



\bibitem {1}
   Busch,P., Lahti,P., Mittelstaedt,P.:
{  Quantum Theory of Measurements}, Springer-Verlag, Berlin,
(1996).



\bibitem {2}
 Jauch,J.M.  { Foundations of Quantum Mechanics},
Addison-Wesly, Reading, (1968).






\bibitem {5}
 Svozil,K.:  { Randomness and undecidability in Physics}.
World Scientific, Singapour, (1993).

\bibitem {6} Schumacher,B.: Quantum Coding.  { Phys. Rev. A}, { \bf {51}}, 2738-2747, (1995).


\bibitem {7}
Nielsen M., Chuang I.: { Quantum Computation and Quantum
Information}
 (Cambridge University Press, Cambridge, U. K., 2000)

\bibitem {8}
Mayburov, S.: Restrictions on Information Transfer in Quantum
Measurements. { Int. J. Quant. Inf.} {\bf {5}}, 279-286, (2007).


 \bibitem {9} Mayburov, S. In: { Frontiers of  Fundamental Physics.}
 AIP conf. proc., vol. 1018, pp. 33-42, Melville, N-Y, (2008).


\bibitem {10} Mayburov,S: Information Transfer in Individual
 Quantum Measurements. Int. J. Quant. Inf. {\bf 9}, 336-343,
 (2011).

\bibitem {3} Breuer,T.: Impossibility of Accurate Self-measurement. { Phil. Sci.} {\bf 62}, 197-214, (1995).

\bibitem {4}
  Breuer,T.: Subjective Decoherence in Quantum Measurements.   { Synthese} {\bf
  {107}}, 1-17,  (1996).









\bibitem {12}
    Zurek,W.: Dynamics of Quantum Measurements. { Phys. Rev. D},  {\bf {26}},
1862-1877, (1982).

\bibitem {11} Grenander,U.: { Pattern Analysis},
Springer-Verlag, Berlin, (1978).

 \bibitem {13}  Elby,A. and Bub,J.: Triorthogonal Uniqness Theorem for Quantum Mechanics.
  { Phys. Rev. A} {\bf 49}, 4213-4216 (1994).


\bibitem {Em4}
  Emch,G.:  { Algebraic Methods in Statistical Physics and
Quantum Mechanics}, Wiley, N-Y, 1972






\bibitem {14}
     Hartle J.B.: Statistical Interpretation of Quantum Mechanics { Amer. J. Phys.}, {\bf  36}, 704-716 (1968).







\end {thebibliography}
\end {document}